\documentclass[conference]{IEEEtran}
\IEEEoverridecommandlockouts
\usepackage{cite}
\usepackage{dblfloatfix}

\usepackage{amsmath,amssymb,amsfonts}
\usepackage{algorithmic}
\usepackage{graphicx}
\usepackage{textcomp}
\usepackage{xcolor}
\def\BibTeX{{\rm B\kern-.05em{\sc i\kern-.025em b}\kern-.08em
    T\kern-.1667em\lower.7ex\hbox{E}\kern-.125emX}}
\begin{document}

\title{Zero-Knowledge Proof (ZKP) Authentication for Offline CBDC Payment System Using IoT Devices\\
}

\author{\IEEEauthorblockN{1\textsuperscript{st} Santanu Mondal}
\IEEEauthorblockA{\textit{Dept. of Computer Science} \\
\textit{Pondicherry University}\\
Puducherry, India \\
iamsantanu21@gmail.com}
\and
\IEEEauthorblockN{2\textsuperscript{nd} T. Chithralekha}
\IEEEauthorblockA{\textit{Dept. of Computer Science} \\
\textit{Pondicherry University}\\
Puducherry, India \\
tchithralekha.csc@pondiuni.ac.in}
}

\maketitle

\begin{abstract}
Central Bank Digital Currency (CBDCs) are becoming a new digital financial tool aimed at financial inclusion, increased monetary stability, and improved efficiency of payment systems, as they are issued by central banks. One of the most important aspects is that the CBDC must offer secure offline payment methods to users, allowing them to retain cash-like access without violating Anti-Money Laundering and Counter-terrorism Financing (AML/CFT) rules. The offline CBDC ecosystems will provide financial inclusion, empower underserved communities, and ensure equitable access to digital payments, even in connectivity-poor remote locations. With the rapid growth of Internet of Things (IoT) devices in our everyday lives, they are capable of performing secure digital transactions. Integrating offline CBDC payment with IoT devices enables seamless, automated payment without internet connectivity. However, IoT devices face special challenges due to their resource-constrained nature. This makes it difficult to include features such as double-spending prevention, privacy preservation, low-computation operation, and digital identity management. 
The work proposes a privacy-preserving offline CBDC model with integrated secure elements (SEs), zero-knowledge proofs (ZKPs), and intermittent synchronisation to conduct offline payments on IoT hardware. The proposed model is based on recent improvements in offline CBDC prototypes, regulations and cryptographic design choices such as hybrid architecture that involves using combination of online and offline payment in IoT devices using secure hardware with lightweight zero-knowledge proof cryptographic algorithm. 

\end{abstract}


\section{Introduction}
The growing pace of digitalisation of payment systems has triggered international curiosity about CBDCs. World over, central banks are investigating the concept of digital money due to a range of motivations that are related to the issue of financial stability and privacy, as well as the lack of scalability of cryptocurrencies, along with private ownership of digital payment infrastructures, monetary autonomy, and sovereignty \cite{1, 6}. According to the Atlantic Council, more than 130 central banks are directly or indirectly involved in developing CBDCs\cite{99}. According to the surveys of the BIS, 2021, 2023, and 2024\cite{6,7},  more than 90 per cent of central banks are undertaking research or carrying out pilots in the field of CBDCs, an involvement that is becoming increasingly popular across the globe \cite{22, 99}.

The capability to support offline payments is the main characteristic of CBDCs so that the services could be accessible in areas with poor connectivity, or when a disaster strikes, or in remote locations. The payment resilience, the minimization of systemic dependencies, and the enhancement of accessibility is also carried out by offline functionality. Nonetheless, the ability to make transactions of offline CBDC then creates a conflict between privacy and regulatory compliance as no online registry will allow transactions to be monitored and screened in real-time \cite{15, 17, 25}.

IoT devices, which include smartcards, wearables, POS terminals and connected sensors, are becoming a central part of the payment infrastructure in the next generation. Their presence in CBDC payment architectures needs to include secure, lightweight, privacy-preserving mechanisms with the ability to work in a limited resources environment. The purpose of the proposal architecture is to create a ZKP-based offline CBDC payment system that could protect transactions by enhancing privacy, compliance, and efficiency on an IoT devices.
\section{Problem Statement}
Offline CBDC transactions enable users to transfer value without network access, giving rise to cash-like anonymity. Nevertheless, it also poses significant enforcement dilemmas to AML/CFT, because now regulators can no longer use continuous identity validation, real-time screening, or an immediate record of transactions as the enforcement mechanisms they employ to stop cybercrime activities among users and businesses in cyberspace\cite{15, 17}. The lack of connectivity poses the dangers of the occurrence of double-spending, hiding an identity, transferring funds, and unrestricted funds flow.
All IoT devices exacerbate these limitations since they have minimal processing power, memory, and battery capacity, making them unsuitable for performing intensive cryptographic tasks. Traditional compliance methods require online ledger checks. Therefore, a new approach is required, one that enables privacy-preserving compliance, lightweight local verification, and secure storage without compromising user anonymity.

% This proposal tackles the problem of designing an IoT-compatible offline CBDC system that uses secure hardware and lightweight Zero-Knowledge Proofs to guarantee privacy, integrity, and regulatory compliance in the absence of network connectivity.

% \section{Research Questions}
% This study is guided by the following research questions:\\
% \textbf{RQ1:} How to securely execute offline CBDC on IoT devices without online consensus?\\
% \textbf{RQ2:} How to prove AML/CFT compliance privately during offline payment?\\
% \textbf{RQ3:} Which ZKP profiles keep verification lightweight on microcontrollers?

\section{Background}
CBDCs are defined as digital forms of sovereign money issued by central banks and denominated in national currency units \cite{9, 14}. They may use distributed ledger technology (DLT) for settlement, though many designs operate independently of blockchain. Retail CBDCs can follow a one-tier model or a two-tier model involving financial institutions \cite{7, 19}.

Before introducing the proposed architecture we first elaborate the fundamental component to establish technical foundation requirement for understanding the proposed system. 

\subsection{Central Bank Digital Currency (CBDC)}
CBDC can be defined as a digital version of currency which is issued by a central bank of a nation. CBDCs are centralised and legally recognised and unlike cryptocurrencies. CBDCs can be majorly categorised as retail CBDC, which is to be used by the public, and wholesale CBDC, which is to be used by interbank settlements and financial institutions.

Also, CBDC systems can be categorised based on an account-based system, where balances are identified by user identities, or a token-based system, where ownership is demonstrated by cryptographic-based digital tokens. The majority of current CBDC architectures have a two-level architecture, in which the central bank is issuing CBDC and the financial intermediaries take care of user onboarding, wallet provisioning and compliance processes. The key design objectives of CBDCs will always be security, scalability, privacy, financial inclusion, and the resilience of the payment system \cite{tang2025taxonomy}. 

\subsection{Offline Payment Systems}
Offline payment systems enable the user to do transactions without having network connectivity. This feature is necessary in rural regions, disaster situations, and unstable internet connection regions. Offline functionality in the context of CBDC allows the usability of cash-like functionality with the advantages of the digital payments.

The most frequently recognized types of offline CBDC design include fully offline, intermittently offline and staged offline system. Complete offline systems enable to carry out offline transactions, however with a high probability of double-spending risk. Intermittently offline systems allow offline payments and periodically synchronize them with an online account, compromising between usability and security. The staged offline systems limit offline use to conditions defined. Among them, intermittently offline systems are generally held as the most viable ones\cite{1}.

\subsection{Double-Spending Problem in Offline Environments}
The issue of double-spending takes place when a digital value is spent twice or more. This issue is avoided in online systems by performing constant verification of ledgers. Offline environments, however, do not have instant access to the central ledger and hence the detection of double-spending is difficult.

Offline CBDC systems are vulnerable to this risk, which can be reduced through device-level controls in the form of secure element, the use of storage resistant to tampering, and cryptographic proofs. These make sure that once a value is used, it cannot be reused even in the case of no connection\cite{}.

\subsection{Privacy and Regulatory Compliance (AML/CFT)}
The financial systems are expected to complay with Anti-Money laundering (AML) and Counter-Terrorism Financing (CFT) which demand monitoring of transactions and restrictions on spending. Meanwhile, the users demand privacy just like physical cash.

The offline CBDC systems cause friction between the goals since the transactions cannot be tracked in real time. Consequently, modern designs aim  to fulfill privacy-compliant requirements, in which the end-users are able to demonstrate compliance to the regulations without disclosing identity or record of operations\cite{michalopoulos2024compliance}.

\subsection{Secure Hardware Technologies}
Hardware components are important in offline CBDC systems. Secure Elements (SE) are cryptography keys that are stored in tamper-resistant chips to perform sensitive operations and cryptography keys securely. Trusted Execution Environments (TEE) offer secure computation in isolated computation units of processors. These technologies allow key extraction, set spending limits and transaction counters, allowing safe offline operation\cite{3}.

\subsection{Device-to-Device Communication}
Offline CBDC transaction between devices rely on short range communication technology. NFC can be used to support contactless communication at distances of a few centimeters, and Bluetooth Low Energy (BLE) has the capability of providing low-power wireless communication over short ranges.
These technologies allows peer-to-peer communication that do not require an internet connection and can be used in the IoT\cite{5}.

\subsection{Zero-Knowledge Proofs (ZKP)}
Zero-Knowledge Proofs enable the prover to prove that he knows a statement without disclosing the knowledge. In CBDC, ZKPs allow a user to demonstrate that he/she has enough money and meets regulative limits without revealing their identity or transactions details.
ZKPs hence present one of the main building blocks to privacy-preserving offline payments\cite{3}.

\section{Motivation}
The proposed work is motivated by the three fundamental requirement that must be satisfied for deploying practical offline CBDC system in IoT environment.
\subsection{Privacy-Preserving} CBDC payments made offline need to be able to offer privacy similar to cash, but comply with the requirements of AML/CFT. According to the fact that real-time observation cannot be conducted in the offline environment, there is the necessity of having means which can demonstrate the regulations adherence without the disclosure of user identities or transactions.
\subsection{Secure IoT Integration}Offline CBDC transactions must offer the privacy like cash, but meet the requirements of AML/CFT. Offline environments cannot support real-time monitoring, and thus, there is the need to come up with mechanisms that can demonstrate compliance to regulations without disclosing user identities or transactions.
\subsection{Lightweight On-Device Verification}IoT devices has small computational and memory capabilities. This means that offline transactions have to be authenticated locally with light weight cryptographic algorithms that are high in security and low in overhead.
\\

This challenges motivate to explore the development of a hybrid offline CBDC architecture that consist of Secure Elements with optimized protocols of ZKP and intermittent synchronization. Such an approach seek to ensure that user retain strong privacy guarantees, regulator maintain verifiable oversight. and IoT devices can execute cryptographic operations within their hardware constraints. Ultimately, the goal is to enable a scalable , privacy-conscious and complying offline CBDC system that seamlessly operate across mobile device, wearable, and embossed IoT plate from.

\section{Literature Survey}

CBDC studies focus on the three key pillars namely, the offline transaction model, trusted hardware implemented on local enforcement, and privacy-preserving cryptographic solutions that allow delivering Anti-Money Laundering (AML) and Counter-Terrorism Financing (CFT) compliance functionality even in the absence of constant connectivity.

Refinements of these elements in the offline digital currency setting have been presented by recent works of central banks and academic scholars in a manner that allows balancing privacy, auditability, and system resilience.

In illustration of the key considerations of central banks and policy frameworks, a central bank CBDC should be able to maintain the accessibility level of cash and heighten its soundness, as well as uphold the requirements of AML/CFT.
\subsection{Privacy-Preserving Compliance in Offline CBDC}
The article by Michalopoulos \textit{et al.} \cite{1} offers a detailed taxonomy of privacy and compliance architecture systems of an offline CBDC, and classifies the systems into full offline, intermittently offline, and hybrid architecture. In their analysis, hybrid designs, a combination of secure hardware and cryptographic privacy proofs are a practice that is a feasible tradeoff between regulation enforcement and anonymity. This literature forms the regulatory and architectural basis of privacy saving but compatible offline payment systems.
\subsection{Trusted Hardware-Based Offline Enforcement}
It is upon this base that Michalopoulos\textit{et al.} \cite{2} put forward a Secure Element (SE)-based offline CBDC prototype to illustrate that tamper-resistant hardware counters, digital certificates and attestation capabilities could be used to achieve resilient offline transactions. Their system ensures high device-level security and low-latency and it is mainly based on hardware attestations and single-device functionality. They also define the necessity to have zero-knowledge proofs (ZKPs) incorporated in a way that will allow compliance to be verified, but user privacy can be maintained. This prototype is further developed in the current work involving SE-based key protection with an addition of Trusted Execution Environments (TEE) to generate collaborated proofs in order to implement multi device sub-wallets providing a more scalable and privacy-aware deployment of IoT.

In addition to these contributions, Beer \textit{et al.}\cite{3} presented a regulated offline CBDC, called \textit{PayOff} \cite{3}, incorporating ZKPs in the account-state transitions, which allows the end-users to attest to compliance with spending limits and enforce sanctions against screening without revealing identities and supported with serial numbers and nullifiers to detect double-spending at reconciliation. The selective auditability that the protocol provides is also based on the court-approved de-anonymization. This architecture defines the methodology of our system that modifies its privacy-compliant model of reconciliation to the resource-constrained devices of an IoT system with lightweight proof generation and device-to-device offline communication.
\subsection{Lightweight Cryptography and IoT-Oriented ZKP Systems}
Although it is named \textit{PayOff} it concentrates on mobile or POS-like hardware, later it has been determined that it is indeed possible to implement ZKPs on constrained IoT platforms. Ramezan and Meamari \cite{4} propose Bulletproofs + and Halo2 modern proof systems prove that can be effectively run on low-power devices. Bulletproofs+ attains transparent range proofs in the size of microcontrollers, and Plonkish/Halo2 achieves much smaller proofs and seconds of verification on the smartphone. This is a reason why we are dual profiled ZKP with Bulletproofs+ on ultra-light devices and Halo 2 on devices with greater capacity to realize scaling of privacy protection across heterogeneous IoT wallets.

Pocher and Zichichi \cite{5} consider the CBDC use in consumer-IoT, and machine-to-machine (M2M) scenarios wherein they discuss the issues of lightweight communication (NFC/BLE) and system interoperability. They state that constrained-device payment figures are viable and that value distribution is required to be efficient amongst the IoT form factors. Our architecture builds this paradigm by a multi-device sub-wallet mechanism, the each device of the IoT implements the spending and compliance limits and rules locally, through the use of SE counters and ZKPs, in order to provide privacy preserving offline interoperability.
\\

Out of these preceding studies, the given methodology brings the problem-solving frameworks of compliance-conserving reconciliation model of \textit{PayOff} \cite{3}, presented by Michalopoulos \textit{et al.} \cite{2} in the article, the security of hardware enforcement concepts of Michalopoulos et al. in the article, the regulatory taxonomy in the article \cite{1}, the lightweight cryptographic optimization of zk-IoT \cite{4} in the article, and the IoT interoperability view of Pocher and Zichichi in the article\cite{5}. They jointly constitute an informative design of a lightweight, IoT-enabled offline CBDC architecture, which has a privacy-preserving AML/CFT existence by locating on a hybrid SE,TEE proving architecture and selective auditability of a threshold.

\section{Research Gap}
Existing literature on the offline CBDC systems has provided privacy preservation, hardware-assistant security, and regulatory compliance through Secure Elements, Trusted Execution environments, and cryptographic techniques. Privacy-preserving compliance verification has also been suggested to be done with Zero-Knowledge Proofs (ZKPs).

However, there are still a number of significant gaps. First, offline CBDC designs continue to be more or less periodic online validating or centralized reconciling, limiting their ability to support fully autonomous offline transactions without online consensus. Second, the existing ZKP-based compliance solutions target primarily smartphone-based or POS devices, and they lack appropriate attention to the resource constant IoT devices. Third, despite the presence of lightweight ZKP frameworks, the recommended way of selecting and mapping suitable ZKP schemes to the heterogeneous IoT platforms while maintain security and compliance are still lacking.

The existing gaps demonstrate that the need for an IoT-compatible offline CBDC architecture. That support offline payment and allow  privacy preserving AML/CFT compliance by employing lightweight ZKP protocols compatible with resource constant devices.
\\This research is guided by the following research questions:\\
\textbf{RQ1:} How to securely execute offline CBDC on IoT devices without online consensus?\\
\textbf{RQ2:} How to prove AML/CFT compliance privately during offline payment?\\
\textbf{RQ3:} Which ZKP profiles keep verification lightweight on microcontrollers?

\section{Proposed Work}
The three research questions that are identified in this study will be answered in the proposed work. 
First, it allows safe implementation of the offline CBDC transactions on the IoT-based devices without necessarily maintaining online consensus by relying on Secure Elements and local enforcement. 
Second, it achive privicy preserving AML/CFT compliances by using Zero-Knowledge Proofs where the user is able to demonstrate regulatory compliance without disclosing sensitive data. 
Third, it implements lightweight ZKP protocol that can be used on resource-constrained IoT devices to have the most efficient on-device verification. 
Based on this principles, a hybrid offline CBDC architecture is presented.
\subsection{System Architecture}

\begin{figure*}[!h]
    \centering
    \includegraphics[width=0.57\linewidth]{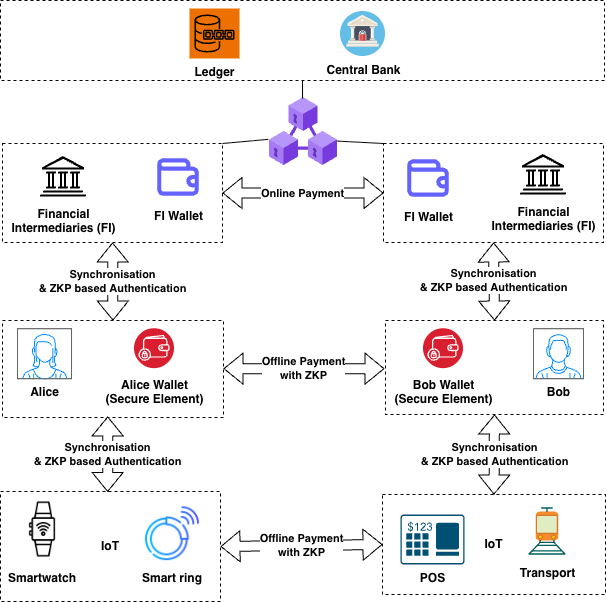}
    \caption{High-level architecture of the proposed offline CBDC system integrating ZKP-based authentication across online and offline payment modes.}
    \label{fig:architecture}
\end{figure*}

Figure~\ref{fig:architecture} illustrates the high level architecture of  the proposed IoT enable offline CBDC system. The architecture is based on a two-tier model of the central bank and distributed ledger infrastructure at the core, financial intermediaries (FIs) to provide wallet provisioning and wallet synchronization, and end-user will have main wallet in mobile devices and with connected sub wallet in IoT devices that have Secure Elements (SEs).

The issuance and the circulation of CBDC in the online mode will be conducted by the central bank, whereas the identity verification and wallet creation are carried out by FIs. At this stage, cryptographic credentials and spending policies to every user wallet and are safely stored within the SE. SE secures the private keys and keeps the monotonic counters to ensure that even during the offline double-spending can be prevent and enforce spending limits.

The system adopts a hierarchical wallet structure composed of a main wallet and multiple sub-wallets. The primary wallet will be deployed on the mobile device (e.g., smartphone) of a user, and will serve as the master store of identity binding, credential management and allocation of values. The primary wallet interacts with finance intermediaries via the online mode to receive CBDC, issue wallet certificates, as well as match transaction records.

IoT devices include wearables, POS terminals, and smart objects that accommodate a number of sub-wallets. Every sub-wallet holds an SE which holds cryptographic keys, fragment of balances and counters used in spending. The central wallet will be in a position to release small portions of CBDC to these sub-wallets to allow them to spend money on their own but impose a limit per device.

Where Trusted Execution Environment (TEE) is present on the mobile devices, it is associated with the SE in helping with sensitive tasks like generation of proofs and enforcement policies. Such hybrid SE-TEE design allows the use of a scalable multi devices with high isolation of secrets.

Peer-to-peer payments in the offline mode take place between sub-wallet to sub-wallet or between a main wallet and a sub-wallet through either near-field communication (NFC) or Bluetooth Low Energy (BLE). The payer generates a Zero-Knowledge Proof (ZKP) proving that they meet four important properties like: (1) possession of relevant funds, (2) transfer value validity, (3) anti-money laundering (AML) and counter-terrorism financing (CFT) limitations and (4) irrevocability of credentials. The payee verify this ZKP proof locally and then the transaction process start executing. All offline transactions are recorded in secure local logs protected by the SE. When the connection is reestablished, such logs are uploaded to be reconciled, detect dual spends, and ensure auditability as authorised by the regulators.main A detailed proposed system work flow is illustrated in Figure \ref{fig:flow}.

\subsection{Transaction Workflow}
\begin{figure*}[!h]
    \centering
    \includegraphics[width=0.7\linewidth]{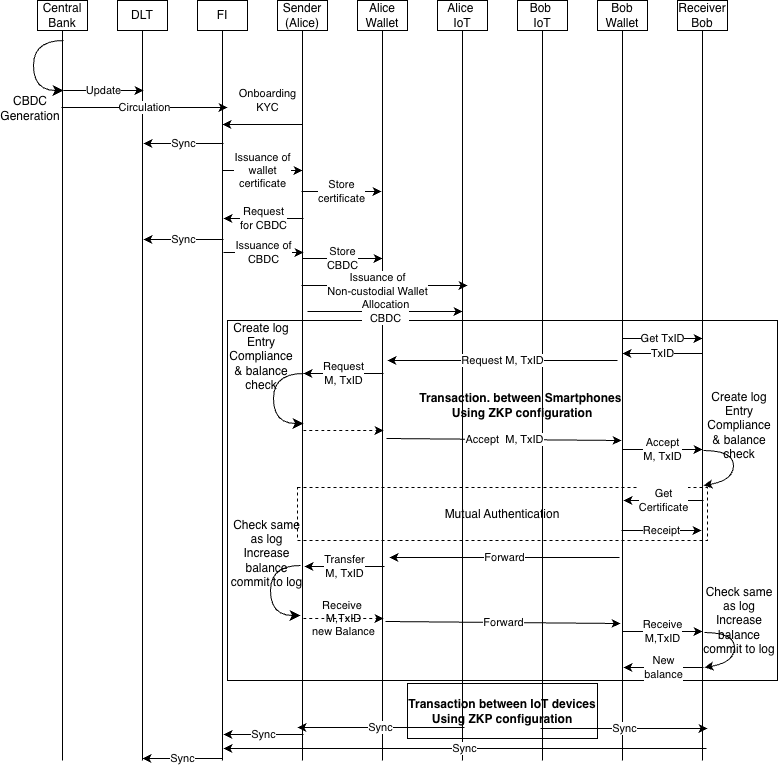}
    \caption{End-to-end workflow of the proposed offline CBDC architecture.}
    \label{fig:flow}
\end{figure*}
Figure~\ref{fig:flow} illustrates the entire flow of a CBDC transaction including wallet on-boarding, value allocation, offline payment exchanges, and offline payment settlement and post transaction synchronization.

First, the Central Bank issues CBDC and makes updates on the circulation details on the Distributed Ledger (DLT). The periodical update of the ledger by Financial Intermediaries (FIs) is done in order to retrieve the current state. A user, Alice getting on boarded by the FI through Know Your Customer (KYC) processes to gt a wallet fpr CBDC. The successful on-boarding consists of the issuance of a wallet certificate by the FI to the primary wallet of Alice mobile device, in which the certificate and cryptographic credentials are stored securely in the Secure Element (SE).

The primary wallet of Alice requests to the FI are for CBDC. When approved, CBDC becomes issued and deposited in the main wallet in the form of non-custodial wallet balance. The primary wallet can then assign some part of its balance to an IoT sub-wallet of Alice where it can be used offline.

In offline payment, an IoT device of Alice will form a transaction request with the message. $M$ (transactions metadata) and $TxID$ (transaction identifier). The device produces a Zero-Knowledge Proof (ZKP) certifying that there is enough balance in it, the amount of the transfer is not invalid, and the spending policies on AML/CFT are met. The request $(M,TxID)$ is transmitted to the IoT device of Bob either with NFC or BLE.

The IoT device of Bob validates the received ZKP and in case it is valid then he accepts the request to perform the transaction. The authentication is done between the two IoT devices between Alice and Bob. The allowed transaction is sent to the wallet of Bob, which retrieves the respective certificate hence creating the receipt. The two devices update their balances on the local level and make secure log entries within their SEs.

After the transaction Bob wallet reflects new balance while Alice wallet reflects the reduced balance. At this phase, there is no online connection and allows peer-to-peer offline payments in the form of payment.

Upon available of network connectivity, the wallet of Alice, and that of Bob, and their respective IoT, update their local logs with the FI. The FI checks logs, eliminates any potential attempts at a double-spending, and reconciles the modified balances with the DLT. This interruptive synchronization, makes it consistent globally and auditable by the regulatory bodies without violating privacy of the users when offline.

\section{Methodology}
The following components will be contained in the methodological approach:
\begin{itemize}
    \item \textbf{Literature and Standards Review:} An analysis of the regulatory frameworks of CBDC (such as BIS, FSB, and FATF), offline design (e.g., Project Polaris), and cryptographic models (such as PayOff and zk-IoT) is undertaken to make sure that they are compatible with the current industry practices and regulatory requirement.
    \item \textbf{System Design:} The hybrid offline CBDC system is designed, which assumes the usage of SE-based wallet storage, ZKP protocols, and intermittent synchronization logic to match the limitations of the IoT devices.
    \item \textbf{ZKP Selection \& Optimization}State-of-the-art ZKP systems will be evaluated, such as Bulletproofs+, Halo2, and Plonkish with respect to few of the previously analyzed aspects like computational overhead, memory footprint, and scalability in validation of a proof under the context of the IoT constraints.
    \item \textbf{IoT Integration} The data communication channels between devices through NFC are planned, and the process of transferring funds among multiple devices as a sub-wallet is designed to address a normal and safe payment in an offline setting.
    \item \textbf{Security \& Compliance Modeling} The rules of AML/CFT are formalized into proof circuits and spending policies that are implemented using SE counter mechanisms, thus making sure that people comply with regulations in the cryptographic infrastructure.

    \item \textbf{Prototype Implementation}The first prototype will be built on simulation platforms to test the measures of performance indicators like latency, verification time and the storage use thus giving empirical validation of the suggested architecture. Upon successful validation, the implementation will be extended to real IoT devices for performance assessment.
    \item \textbf{Evaluation} The prototype will be then evaluated based on performance matrices, resource utilization matrices, security and privacy matrices, scalability matrices.

\end{itemize}

\section{Conclusion and Future Work}
This paper propose a hybrid SE and ZKP-enabled offline CBDC architecture for IoT environment with the goal to achive privacy, security, operational efficiency, and compliance with regulations. The design addresses key issues arising from intermittent connectivity such as double spending prevention, credential management, and local verification while supporting AML/CFT requirements without using sensitive user data.

The proposed framework, which consists of a combination of lightweight ZKP schemes and tamper-resistant SE that can be periodically synchronized. This allows performing cash-like offline payments with selective auditability (i.e.; Controlled disclosure of transaction information to authorized auditor while preserving users anonymity). The system can be used in heterogeneous IoT environments and resource-constrained IoT systems as it supports multi-device sub-wallet and the NFC/BLE-based communication.

The further work will be based on the implementation of a complete prototype and testing it with different types of IoT, including wearables, smartcards, and embedded entities. The performance will be tested based on the key performance, resource, security, and reliability indicators, such as the proof generation and verification time, the time spent on transactions offline, the delayed to synchronize, the memory and storage overhead, the cost of computation, the size of the proof, and the rate of a transaction, the integrity of transactions, privacy maintenance, the throughput of transactions, and after synchronization reliability. Future research directions can be seen in the combination of threshold signatures and secure multiparty computation as further ways to get an even higher degree of scalability and robustness.

Moreover, the prototype 
% \begin{figure}[htbp]
% \centerline{\includegraphics{fig1.png}}
% \caption{Example of a figure caption.}
% \label{fig}
% \end{figure}

% \begin{table}[htbp]
% \caption{Table Type Styles}
% \begin{center}
% \begin{tabular}{|c|c|c|c|}
% \hline
% \textbf{Table}&\multicolumn{3}{|c|}{\textbf{Table Column Head}} \\
% \cline{2-4} 
% \textbf{Head} & \textbf{\textit{Table column subhead}}& \textbf{\textit{Subhead}}& \textbf{\textit{Subhead}} \\
% \hline
% copy& More table copy$^{\mathrm{a}}$& &  \\
% \hline
% \multicolumn{4}{l}{$^{\mathrm{a}}$Sample of a Table footnote.}
% \end{tabular}
% \label{tab1}
% \end{center}
% \end{table}

\bibliographystyle{IEEEtran}
\bibliography{references}

@techreport{6,
  title={Project Polaris: A handbook for offline payments with CBDC},
  author={Bank for International Settlements},
  institution={BIS},
  year={2023},
  url={https://www.bis.org/publ/othp64.htm}
}

@techreport{7,
  title={Central bank digital currencies: foundational principles and core features},
  author={Bank for International Settlements},
  institution={BIS},
  year={2020},
  url={https://www.bis.org/publ/othp33.htm}
}

@article{9,
  title={Cashless society–the future of money or a utopia?},
  author={Fabris, Nikola},
  journal={Journal of Central Banking Theory and Practice},
  volume={8},
  number={1},
  pages={53--66},
  year={2019}
}

@article{14,
  title={PUF-based digital money with offline transfers},
  author={Bean, Ben and Minwalla, C and Tsiropoulou, EE and Plusquellic, J},
  journal={Journal of Emerging Technologies in Computing Systems},
  year={2024}
}

@techreport{15,
  title={Ready, steady, go? Results of the third BIS survey on central bank digital currency},
  author={Boar, Codruta and Wehrli, Andreas},
  institution={Bank for International Settlements},
  year={2021}
}

@article{17,
  title={Offline payments with CBDC: Challenges and design considerations},
  author={Christodorescu, Mihai and Gu, Wei-Cheng and Kumaresan, R.},
  journal={Microsoft Research Technical Report},
  year={2020}
}

@article{19,
  title={Cross-border payment system modernization and CBDCs},
  author={Auer, Raphael and Claessens, Stijn},
  journal={BIS Papers},
  year={2022}
}

@article{22,
  title={Central Bank Digital Currencies: Economic Considerations and Financial Stability Implications},
  author={Bindseil, Ulrich and Panetta, Fabio},
  journal={ECB Occasional Paper Series},
  year={2023}
}

@article{25,
  title={Balancing privacy and compliance in digital currency architecture},
  author={Kahn, Charles and Roberds, William},
  journal={Journal of Monetary Economics},
  year={2021}
}

@misc{99,
  title={Central Bank Digital Currency Tracker},
  author={Atlantic Council},
  year={2024},
  url={https://www.atlanticcouncil.org/cbdctracker/}
}

@article{1,
  author  = {Panagiotis Michalopoulos and Odunayo Olowookere and Nadia Pocher and Johannes Sedlmeir and Andreas Veneris and Poonam Puri},
  title   = {Privacy and Compliance Design Options in Offline Central Bank Digital Currencies},
  journal = {IEEE Transactions on Network and Service Management},
  year    = {2025},
  volume  = {22},
  number  = {5},
  pages   = {3748--3762}
}

@article{2,
  author    = {Panagiotis Michalopoulos and Anthony Mack and Cameron Clark and Linus Chen and Johannes Sedlmeir and Andreas Veneris},
  title     = {Balancing Compliance and Privacy in Offline CBDC Transactions Using a Secure Element-based System},
  journal   = {arXiv preprint arXiv:2509.25469},
  year      = {2025}
}

@article{3,
  author  = {Carolin Beer and Sheila Zingg and Kari Kostiainen and Karl W{\"u}st and Vedran Capkun and Srdjan Capkun},
  title   = {PayOff: A Regulated Central Bank Digital Currency with Private Offline Payments},
  journal = {arXiv preprint arXiv:2408.06956},
  year    = {2024}
}

@inproceedings{4,
  author    = {Gholamreza Ramezan and Ehsan Meamari},
  title     = {zk-IoT: Securing the Internet of Things with Zero-Knowledge Proofs on Blockchain Platforms},
  booktitle = {2024 IEEE International Conference on Blockchain and Cryptocurrency (ICBC)},
  address   = {Dublin, Ireland},
  pages     = {1--7},
  year      = {2024},
  doi       = {10.1109/ICBC59979.2024.10634342}
}

@inproceedings{5,
  author    = {Nadia Pocher and Mirko Zichichi},
  title     = {Towards {CBDC}-based Machine-to-Machine Payments in Consumer {IoT}},
  booktitle = {Proceedings of the 37th ACM/SIGAPP Symposium on Applied Computing (SAC '22)},
  address   = {New York, NY, USA},
  publisher = {ACM},
  year      = {2022},
  doi       = {10.1145/3477314.3507078}
}

@misc{tang2025taxonomy,
  author = {Qian Tang and Lu Si},
  title  = {CBDC System Design Taxonomy and Ecosystem: A Survey of 2024--2025 Literature},
  year   = {2025},
  note   = {Preprint}
}

@inproceedings{michalopoulos2024compliance,
  title={Compliance design options for offline CBDCs: Balancing privacy and AML/CFT},
  author={Michalopoulos, Panagiotis and Olowookere, Odunayo and Pocher, Nadia and Sedlmeir, Johannes and Veneris, Andreas and Puri, Poonam},
  booktitle={2024 IEEE International Conference on Blockchain and Cryptocurrency (ICBC)},
  pages={307--315},
  year={2024},
  organization={IEEE}
}
\end{document}